\documentclass[10pt,twocolumn]{ICCAS}
 
\usepackage{tabu}
\usepackage{diagbox}
\usepackage{cite}
\usepackage{array}
\usepackage{comment}
\usepackage{anyfontsize}
\usepackage{mathtools}
\usepackage{amsmath, dsfont, bm}
\usepackage{dsfont}
\usepackage{algorithm}
\usepackage{algpseudocode}

\begin{document}

\title{Evaluating Particle Filtering for RSS-Based Target Localization under Varying Noise Levels and Sensor Geometries}
\author{Halim Lee${}^{1}$, Jongmin Park${}^{1}$, and Kwansik Park${}^{2*}$ }

\affils{ ${}^{1}$School of Integrated Technology, Yonsei University, \\
Incheon, 21983, Korea (halim.lee, jm97@yonsei.ac.kr) \\
${}^{2}$Korea Aerospace Research Institute, \\
Daejeon, 34133, Korea (kspark6469@kari.re.kr) \\
{\small${}^{*}$ Corresponding author}}

\abstract{
Target localization is a critical task in various applications, such as search and rescue, surveillance, and wireless sensor networks. When a target emits a radio frequency (RF) signal, spatially distributed sensors can collect signal measurements to estimate the target's location. Among various measurement modalities, received signal strength (RSS) is particularly attractive due to its low cost, low power consumption, and ease of deployment. While particle filtering has previously been applied to RSS-based target localization, few studies have systematically analyzed its performance under varying sensor geometries and RSS noise levels. This paper addresses this gap by designing and evaluating a particle filtering algorithm for localizing a stationary target. The proposed method is compared with a conventional RSS-based trilateration approach across different sensor configurations and noise conditions. Simulation results indicate that particle filtering provides more accurate target localization than trilateration, particularly in scenarios with unfavorable sensor geometries and high RSS noise.
}


\maketitle


\section{Introduction}
Target localization is a critical task in a wide range of applications, including search and rescue (SAR), surveillance, robotic applications \cite{Moon22:Sample, Li25:Time, Moon22:Fast}, and wireless sensor networks (WSNs). In emergency scenarios where the target (e.g., a person in distress) is unable to provide location information using the global navigation satellite system (GNSS) \cite{Kim25:Set, Zhu18:GNSS, Lee22:Urban, Lee22:Optimal, Kim23:Machine, Jeong24:Quantum, Kim23:Low, Lee22:SFOL, Jia21:Ground, Park25:Toward, Lee25:A, Kim24:Performance, Lee24:A, Chiou08:Performance, Sun21:Markov, Chen10:Real, DeLorenzo10:WAAS/L5, Kim23:Single, Lee23:Seamless}, an alternative is to estimate the target's position directly from signal measurements collected by nearby sensors or vehicles \cite{Lee25:Reducing, Moon24:HELPS, Lee23:Performance_Comparison}.

In such cases, multiple spatially distributed sensors can obtain signal measurements from the radio frequency (RF) signal transmitted by the target. These measurements may include time-of-arrival (ToA) \cite{Nguyen16:Optimal, Wang13:Convex, Lee23:Nonlinear}, time-difference-of-arrival (TDoA) \cite{Kim22:First, Son18:Novel, Son19:Universal, Son24:eLoran, Rhee21:Enhanced, Kang25:Enhancing}, and received signal strength (RSS) \cite{Lee22:Evaluation, Zanella16:Best, Lee22:Performance}. Among these modalities, RSS is particularly attractive due to its low hardware cost, low energy consumption, and ease of deployment.

Despite its practicality, RSS-based target localization presents notable challenges. First, when formulating localization as a maximum likelihood estimation (MLE) problem using RSS measurements, the resulting optimization problem lacks a closed-form solution due to the nonlinearity of the signal propagation model \cite{Lee23:Performance_Comparison, Vaghefi12:Cooperative, Jeong20:RSS}. Second, the estimation performance is often sensitive to the geometric configuration of the sensors and measurement noise.

To address the first challenge, various algorithms have been proposed, including convex relaxation methods \cite{Vaghefi12:Cooperative, Ouyang10:Received}, heuristic approaches such as genetic algorithms \cite{Lee21:Genetic, Ren20:Rssi}, and probabilistic inference techniques \cite{Wagle10:Particle, Ullah19:Localization}. Among the probabilistic methods, particle filtering has been considered a promising approach for RSS-based localization \cite{Wagle10:Particle, Ullah19:Localization, Tomic17:Bayesian}. It enables the approximation of a posterior distribution over the target location, rather than producing a single point estimate, and can accommodate nonlinear and non-Gaussian measurement models.

While prior studies have applied particle filtering to RSS-based target localization problems, the influence of measurement noise and sensor geometry on its performance has not been thoroughly analyzed. To bridge this gap, this paper investigates the performance of particle filtering for stationary target localization under different sensor deployment geometries and varying levels of RSS measurement noise.

\section{Particle Filtering with RSS Measurements}
We consider a stationary target located at an unknown position $\mathbf{x} \in {\mathds{R}}^{\mathrm{2}}$, and $M$ sensor nodes at known locations $\mathbf{s}_{\mathit{m}} \in {\mathds{R}}^{\mathrm{2}}$. Each sensor records an RSS value modeled as \cite{Vaghefi12:Cooperative, Li07:Collaborative}:
\begin{equation}
\hat{P}_m = P_0 - 10\beta \log_{10}(\|\mathbf{x} - \mathbf{s}_{\mathit{m}}\|) + \mathit{v}_{\mathit{m}},
\end{equation}
where $P_0$ is the reference RSS (at 1 m), $\beta$ is the path loss exponent, and $v_m\!\sim\!\mathbf{N}(\mathrm{0},\mathit{\sigma}^{\mathrm{2}})$ denotes zero-mean Gaussian noise in the RSS measurements. This study assumes that the values of $P_0$ and $\beta$ are available a priori or can be estimated using well-known techniques proposed in prior studies \cite{Vaghefi12:Cooperative, Jeong20:RSS}.

The proposed particle filtering algorithm for estimating the target's position is detailed in Algorithm~\ref{alg:pf}. The particle filter begins with an initialization step, in which $N$ particles $\mathbf{p}^{(\mathit{i})} \in {\mathds{R}}^{\mathrm{2}}$ are uniformly generated over the search area. This initialization is performed only once at the beginning. As RSS measurements are collected over time, the algorithm repeatedly performs the weighting, resampling, and estimation steps. 

In the weighting step, each particle is assigned a weight $w^{(i)}$ based on the likelihood, defined as
\begin{equation}\label{eq:weight}
w^{(i)} \propto \exp\left( -\frac{1}{2\sigma^2} \sum_{m=1}^M \left( \hat{P}_m - P_m^{(i)}\right)^2 \right),
\end{equation}
\begin{equation}\label{eq:particleRSS}
P_m^{(i)} = P_0 - 10\beta \log_{10}\left(\|\mathbf{p}^{(\mathit{i})} - \mathbf{s}_{\mathit{m}}\| + \mathit{\varepsilon} \right), 
\end{equation}
where $\varepsilon$ is a small positive constant added to avoid the singularity of the logarithmic function when the distance approaches zero.

In the resampling step, particles are resampled according to their normalized weights. In Algorithm~\ref{alg:pf}, $\rho$ denotes the resampling ratio that controls the proportion of particles retained based on their importance weights. Finally, in the estimation step, the target's position is given by the mean of the resampled particles.

\begin{algorithm}[tbp!] 
\caption{Particle Filter for RSS-Based Localization} \label{alg:pf}
\begin{algorithmic}[1] 
\Require Sensor positions $\{\mathbf{s}_{\mathit{m}}\}_{\mathit{m}=1}^{\mathit{M}}$, RSS measurements $\{\hat{P}_m\}_{m=1}^M$, $P_0$, $\beta$, $\sigma$, search area
\Ensure Estimated target position $\hat{\mathbf{x}}$

\State Initialize $N$ particles $\{\mathbf{p}^{(\mathit{i})}\}_{\mathit{i}=1}^{\mathit{N}}$ uniformly in the search area
\For{$i = 1$ to $N$}
    \For{$m = 1$ to $M$}
        \State Compute predicted RSS:
        \[
        \hspace{2.2em} P_m^{(i)} \gets P_0 - 10 \beta \log_{10}(\| \mathbf{p}^{(\mathit{i})} - \mathbf{s}_{\mathit{m}} \| + \varepsilon)
        \]
    \EndFor
    \State Compute importance weight:
    \[
    \hspace{-0.8em} w^{(i)} \gets \exp\left( -\frac{1}{2\sigma^2} \sum_{m=1}^M (\hat{P}_m - P_m^{(i)})^2 \right)
    \]
\EndFor
\State Normalize weights: $w^{(i)} \gets w^{(i)} / \sum_j w^{(j)}$
\State Resample top $\rho \cdot N$ particles based on weights
\State Generate $(1 - \rho) \cdot N$ new particles uniformly
\State Combine resampled and new particles to form updated set
\State Estimate $\hat{\mathbf{x}}$ as the mean of all particles
\State \Return $\hat{\mathbf{x}}$
\end{algorithmic}
\end{algorithm}

\section{Evaluation}
\subsection{Effect of Sensor Placement Configurations}

\begin{figure} [tbp!]
    \centering 
    \includegraphics[width=1\linewidth]{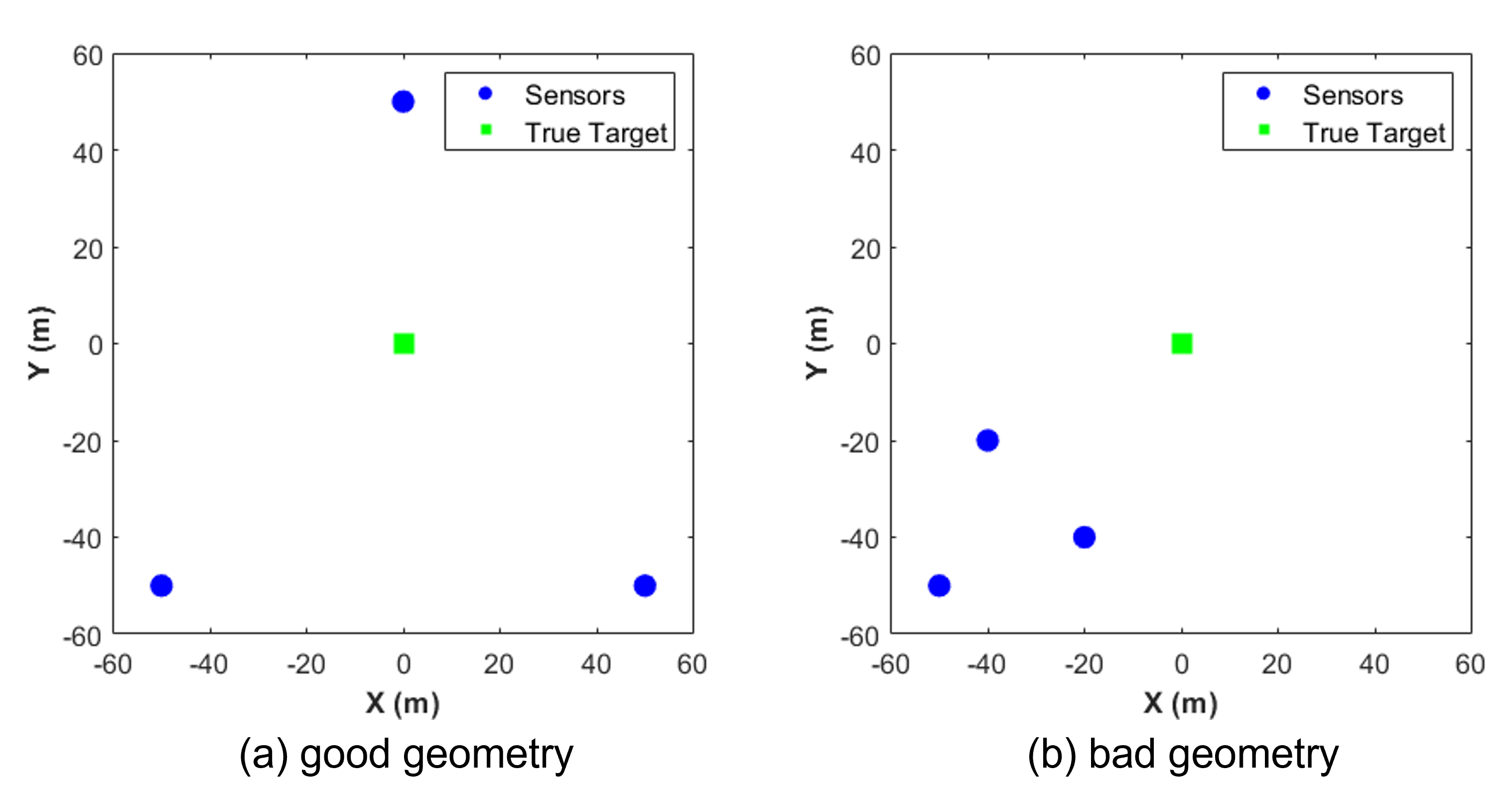}
    \caption{Sensor placement scenarios used for performance evaluation. (a) Good geometry case, where sensors are evenly distributed around the target; (b) Bad geometry case, where sensors are clustered on one side.} 
    \label{fig:setting}
\end{figure}

We first evaluated the proposed particle filter under two sensor deployment scenarios. Fig.~\ref{fig:setting}(a) illustrates the good geometry case, where the sensors are well-distributed around the target. In contrast, Fig.~\ref{fig:setting}(b) shows the bad geometry case, where the sensors are clustered on one side of the target.

For comparison, we implement a basic RSS-based trilateration method. The estimated distance from the RSS measurement is:
\begin{equation}
\hat{d}_m = 10^{\frac{P_0 - \hat{P}_m}{10\beta}},
\end{equation}
and the target position is estimated by solving:
\begin{equation} \label{eq:tril}
\hat{\mathbf{x}} = \arg\min_{\mathbf{x}} \sum_{m=1}^M \left( \|\mathbf{x} - \mathbf{s}_{\mathit{m}}\| - \mathit{\hat{d}_{m}} \right)^2.
\end{equation}

All implementations were developed in MATLAB, and the \texttt{fminsearch} function was used to solve the optimization problem in Eq.~\eqref{eq:tril}. In the simulation, the reference RSS value $P_0$ was set to -30~dB, the path loss exponent $\beta$ was set to 2.5, and the standard deviation of the RSS noise $\sigma$ was set to 5~dB. The search area was defined as a square region ranging from $(-50, -50)$ to $(50, 50)$. The resampling ratio \( \rho \) was set to 0.9.

\begin{figure} [tbp!]
    \centering 
    \includegraphics[width=0.85\linewidth]{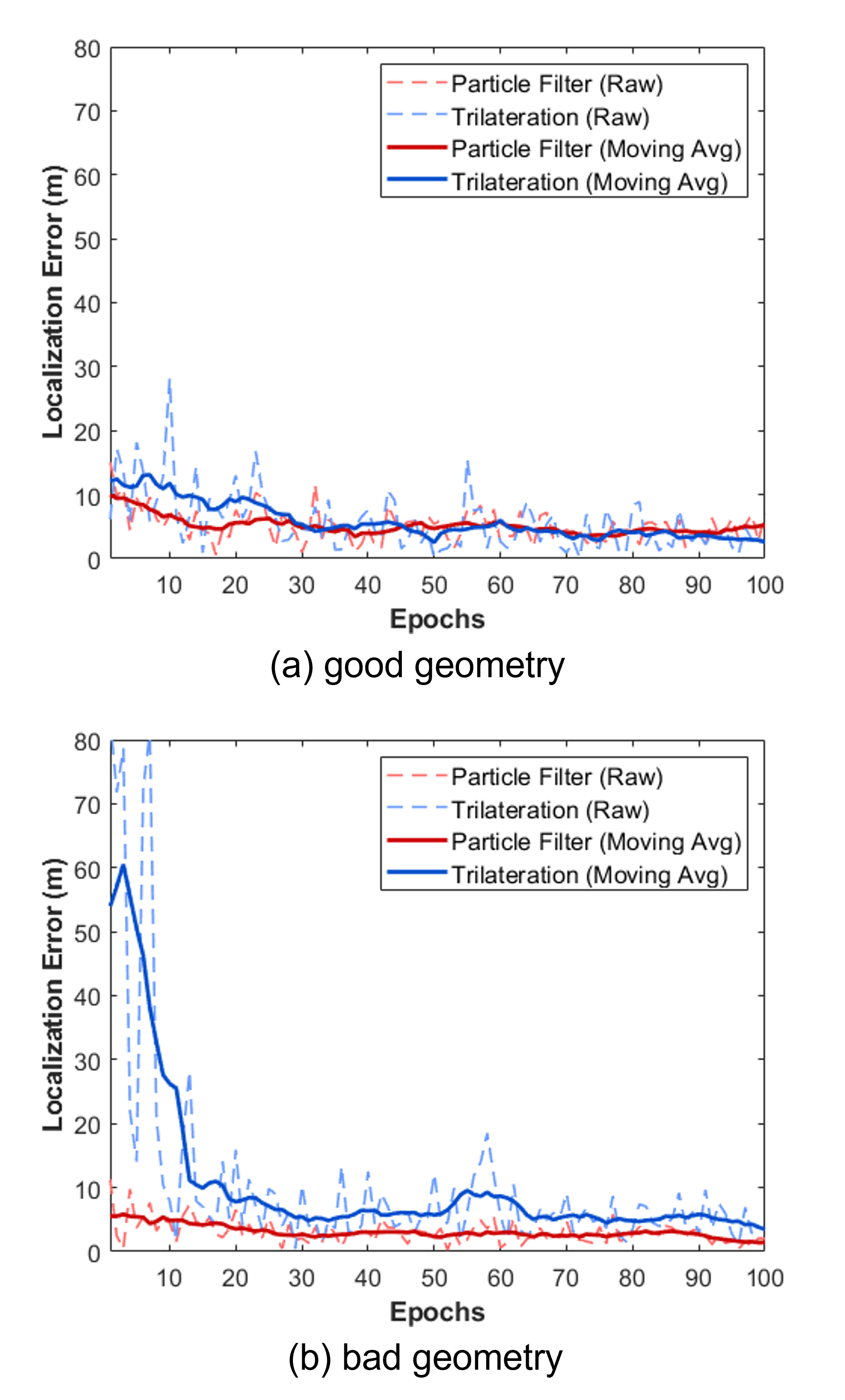}
    \caption{Localization error over time epochs for (a) the good geometry case and (b) the bad geometry case.} 
    \label{fig:result}
\end{figure}

Fig.~\ref{fig:result} presents the localization error over time epochs (i.e., the number of RSS measurements). In the good geometry scenario, both methods achieve comparable performance. However, in the bad geometry scenario, the particle filter performs better than trilateration, which can be attributed to its ability to more effectively handle geometric uncertainty.

\subsection{Effect of RSS Noise Level}

We further compared the performance of the particle filter and RSS-based trilateration under varying levels of RSS measurement noise in the bad geometry scenario shown in Fig.~\ref{fig:setting}(b). In realistic signal reception environments, RSS measurements can be affected by surrounding obstacles, resulting in significant measurement noise \cite{Lee23:Performance_Evaluation}. To evaluate this effect, simulations were conducted for noise levels ranging from 1~dB to 10~dB.

\begin{figure} 
    \centering 
    \includegraphics[width=0.85\linewidth]{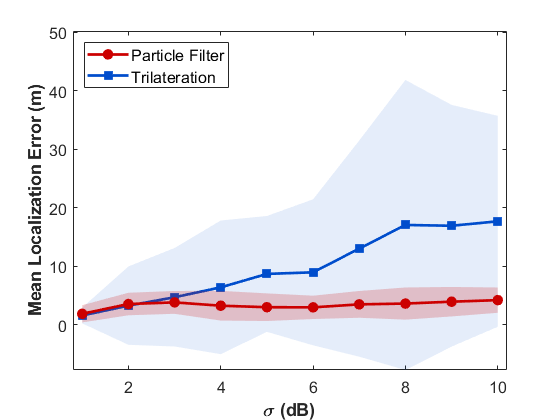}
    \caption{Average localization error over 100 time epochs versus RSS noise level in the bad geometry case.} 
    \label{fig:std_result}
\end{figure}

Fig.~\ref{fig:std_result} illustrates the average localization error over 100 time epochs for each noise level. The solid lines represent the mean localization error, while the shaded areas (with the same color as the corresponding lines) indicate the standard deviation. The results show that the two methods perform similarly at low noise levels (1–2 dB). However, for noise levels of 3~dB and above, the particle filter consistently outperforms trilateration. Furthermore, the particle filter exhibits lower variance across all noise levels, indicating more stable and reliable performance.

\section{Conclusion}
Accurate localization of a target is essential in various applications, particularly in emergency scenarios where GNSS-based location information is not accessible. In such localization scenarios, RSS measurements provide a low-cost and readily available alternative. Although particle filtering has been used in previous studies for RSS-based localization, the effects of sensor geometry and measurement noise on its performance have not been thoroughly investigated. This study fills that gap by systematically evaluating a particle filtering-based localization algorithm and comparing it with a conventional RSS-based trilateration method. The results show that particle filtering offers improved accuracy and stability in challenging conditions, such as clustered sensor deployments and high-noise environments.

\section*{ACKNOWLEDGEMENT}

This work was supported in part by the National Research Foundation of Korea (NRF), funded by the Korean government (Ministry of Science and ICT, MSIT), under Grant RS-2024-00358298; 
in part by the Korea Aerospace Administration (KASA), under Grant RS-2022-NR067078; 
in part by Grant RS-2024-00407003 from the ``Development of Advanced Technology for Terrestrial Radionavigation System'' project, funded by the Ministry of Oceans and Fisheries, Republic of Korea; 
in part by the Unmanned Vehicles Core Technology Research and Development Program through the NRF and the Unmanned Vehicle Advanced Research Center (UVARC), funded by the MSIT, Republic of Korea, under Grant 2020M3C1C1A01086407; 
in part by the Institute of Information \& Communications Technology Planning \& Evaluation (IITP) through the Information Technology Research Center (ITRC) program, funded by the MSIT, under Grant IITP-2025-RS-2024-00437494.

Generative AI (ChatGPT, OpenAI) was used solely to assist with grammar and language improvements during the manuscript preparation process.  
No content, ideas, data, or citations were generated by AI.  
All technical content, methodology, analysis, and conclusions were written and verified solely by the authors.

\bibliographystyle{IEEEtran}
\bibliography{IUS_publications, mybibfile}

\end{document}